\documentclass[doublecol]{epl2}
\usepackage{amsmath}
\usepackage{amssymb}
\usepackage{amsfonts}
\usepackage{graphicx}
\usepackage{dcolumn}
\usepackage{bm}
\usepackage{sidecap}
\usepackage {caption}
\usepackage{float}
\usepackage{parskip}

\title{Balanced condition in networks leads to Weibull statistics}

\author{Sarika Jalan\thanks{sarika@iiti.ac.in} and Sanjiv K. Dwivedi}
\institute{Complex Systems Lab, Indian Institute of Technology Indore,
IET-DAVV Campus Khandwa Road, Indore-452017}


\abstract{
The importance of the balance in inhibitory and excitatory couplings in the brain 
has increasingly been realized. Despite the key role played by 
inhibitory-excitatory couplings in the functioning of brain networks, the impact of 
a balanced condition on the stability properties of underlying networks remains largely
unknown. We investigate properties of the largest eigenvalues 
of networks having such couplings, and find that they follow
completely different statistics when in the balanced situation.
Based on numerical simulations, we demonstrate that the transition from 
Weibull to Fr\'echet via the Gumbel distribution can be controlled by the
variance of the column sum of the adjacency matrix,
which depends monotonically on the denseness of the underlying
network. As a balanced condition is imposed, the largest real part of the eigenvalue 
emulates a transition to the generalized extreme value statistics, independent of the inhibitory connection probability. Furthermore, the transition to the Weibull statistics
and the small-world transition
occur at the same rewiring probability, reflecting a more stable system.
}

\begin{document}
\maketitle 
\section{Introduction}
The largest eigenvalue of network adjacency matrices plays a bridge between dynamical and 
structural properties of an underlying system. For example, the
inverse of the largest eigenvalue of a network characterizes the
threshold for phase transition of the virus spread \cite{Mieghem2009}. Recently
Goltsev et. al. have demonstrated the importance of the largest eigenvalue in determining 
disease spread in complex networks \cite{Mendes2012}. Furthermore, in coupled oscillators the threshold for 
phase transition to synchronized behaviour is determined by the inverse of the largest eigenvalue 
\cite{Restrepo}. The dynamical properties of neurons have been shown to be highly influenced by 
a change in the spectra of underlying synaptic matrices constructed from randomly distributed numbers  
\cite{Sompolinsky}. 
A remarkable, fundamental direction to analyze the stability of ecological systems  
was put forward by May \cite{MayNature1972}, where
the largest real part of the eigenvalues ($R_{\mathrm max}$) establishes a
relationship between the stability and complexity of the underlying
system. Later, the impact of various types of interactions was demonstrated to
deduce stability criteria in terms of $R_{\mathrm max}$ \cite{Allesina}. 
Mathematically, matrices obeying some constraints satisfy the stability criteria
\cite{Quirck1965}, but real-world systems have an underlying interaction matrix that is
too complicated to obey these constraints; hence, the study of fluctuations in 
$R_{\mathrm max}$ is crucial to understanding stability of a
system as well as the stability properties of an individual network 
in that ensemble. Recent efforts in this direction reveal the similarity of the
maximal Lyapunov exponent of synaptic matrices defined for neural networks with their topological complexities
\cite{Gilles}. A very recent work investigates the
statistical properties of random matrices within the framework of extreme value 
theory, thereby providing an estimation about the resolution in complex dynamics for a
finite system size \cite{Luis}.

\subsection{Balanced condition and its role in stability}
The balanced condition in the brain refers to a situation in which for each neuron the weight of 
the inhibitory signal is equal to the excitatory signal \cite{Shadlen,Troyer}. Ref. \cite{Rajan}
demonstrates that this condition 
forces outliers of the spectra to appear inside the bulk, leading
to a stable underlying neural system.
Further analysis of a dynamical model of cortical networks with the balanced condition for various ratios 
of inhibitory and excitatory 
neurons reveals a connection between the spectra of connectivity matrix and the
dynamical response \cite{Non-normal}. Balance between recurrent excitation and inhibition generates stable periods of 
activity\cite{Compte}. There have been several discussions on how synaptic matrices in the brain
achieve the balanced condition; for instance, it has been demonstrated that the balanced condition
in sensory pathways and memory networks is maintained through a plasticity mechanism at inhibitory synapses \cite{Vogels2}.
In addition to the research emphasizing the importance of the balanced condition, there exist
papers discussing the relevance of different ratios of inhibitory and excitatory neurons in brain; for example,
cortical neurons consist of only $20-30\%$ inhibitory neurons \cite{DiverseInhibitory}.

\subsection{Extreme value theory and its relevance}
The extremal eigenvalue statistics are widely used in various 
disciplines of science.  
The generalized extreme value distribution (GEV) is applied to model 
extrema of independent, identically distributed random variables. 
GEV statistics have been realized in many real-world and model systems.
For example, the radius of the bulk of complex eigenvalues of non-Hermitian
random matrices has been shown to follow the Gumbel distribution \cite{Rider}.
Recent research revealing a rich network architecture has given way to the
spectral studies of matrices deviating from a random structure. One of these
studies demonstrates that statistics of largest 
eigenvalue of matrices with entries following the power-law
distribution displays a transition 
from the Fr\'echet to the Tracy-Widom distribution at a threshold governed by the 
power-law exponent \cite{Biroli}.
The statistics of the inverse of the largest eigenvalue for an ensemble of scalefree networks follows the Weibull
distribution \cite{SF_inverse}. Some of the studies pertaining to 
sparse random graphs,  
and gain matrices in the context of brain networks, are shown to deviate from GEV
statistics and follow normal distribution instead \cite{Komlos}.
The statistical properties of $R_{\mathrm max}$ of synaptic matrices capturing
inhibitory and excitatory couplings reveal a 
transition to the extreme value distribution\cite{Extreme}. However, the extreme value 
distribution is not observed for a larger parameter regime, thereby restricting the applicability of 
the results for a more realistic underlying network construction.  

Extreme value statistics for independent,
identically distributed random variables can be formulated entirely in terms of
three universal types of probability functions: the Fr\'echet, Gumbel and
Weibull distributions, also known as GEV statistics depending 
upon whether the tail of the density
is power-law, faster than any power-law, and bounded or unbounded respectively
\cite{book_gev}.
GEV statistics
with a location parameter $\mu$, scale parameter $\sigma$ and shape parameter $\xi$ 
have often been used to model unnormalized data
from a given system.
The probability density function for extreme value statistics
with these parameters is given by \cite{book_gev},
\makeatletter
\def\@eqnnum{{\normalsize \normalcolor (\theequation)}}
 \makeatother
{ \small
\begin{equation}
\rho(x) = \begin{cases} \frac{1}{\sigma}\big[1+\big(\xi\frac{(x-\mu)}{\sigma}\big)^{-1-\frac{1}{\xi}}\big]\exp\big[-\big(1+\big(\xi\frac{(x-\mu)}{\sigma}\big)^{-\frac{1}{\xi}}\big)\big]&\\ 
\hspace*{5.6cm}\mbox{if } \xi\not=0 \\
\frac{1}{\sigma}\exp\big(-\frac{x-\mu}{\sigma}\big)\exp\big[-\exp\big(-\frac{x-\mu}{\sigma}\big)\big] 
~~ \mbox{if } \xi=0. \end{cases}
\label{eq_gev}
\end{equation}
}
Distributions associated with $\xi > 0$ , $= 0$ and $< 0$ are characterized by the Fr\'echet, Gumbel, 
and Weibull distributions respectively. 
   
In this Letter, we investigate the statistical properties of $R_{\mathrm max}$ for 
networks in the balanced condition. The factors affecting the balanced condition 
are  monitored. We witness the Weibull distribution for the strictly balanced condition. 
Observed behaviour is not much affected by the change in underlying architecture; rather 
it depends more on the denseness of connections. We present
results for Erd\"os-R\'enyi random networks, small-world networks and scalefree networks.
\begin{figure}[t]
\centerline{\includegraphics[width=0.9\columnwidth,height=4cm]{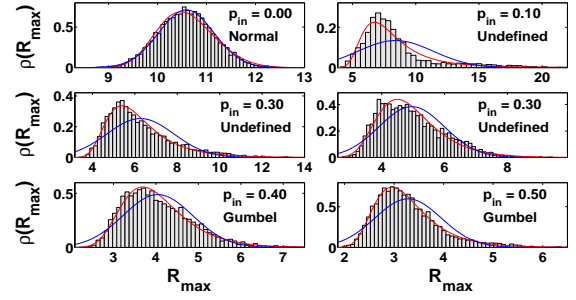}}
\caption{(Colour online) Statistics of $R_{\mathrm max}$ at different values
of  $p_{\mathrm in}$. The histograms are
numerical results; blue and red lines correspond to normal
and GEV fit, respectively. For each case, the balanced matrix
is constructed for network parameters $N=50$ and $<k>=10$.
All plots are obtained for 5000 realizations of the network in that ensemble.}
\label{Fig_ER_Stat}
\end{figure}
\section{Model}
The balanced condition is attained by assigning a fixed weight to inhibitory
and excitatory connections  
in the following manner \cite{Non-normal}.
When a node is defined as inhibitory with probability $p_{\mathrm in}$, the 
corresponding entry in that row of the matrix $A$ is replaced by 
$1-1/p_{\mathrm in}$.
In the matrices constructed as above, most of the column sum would be
fluctuating closely about the zero value for $p_{\mathrm in}$, lying in the vicinity of
0.50. However, for 
lesser values of 
$p_{\mathrm in}$ there may be some columns that have only excitatory connections,
yielding only $zero$ or $+1$ entries, which hinder the achievement of
the balanced condition. 
Furthermore, we achieve a strictly balanced condition by subtracting a constant term
from each non-zero element of a column, which restricts the sum of the 
column entries to a zero value. The strictly balanced condition 
is defined to resolve situations where the arrangement of inhibitory and excitatory
couplings leads to a fluctuation around the zero value for the column sum,
even after imposing
the balanced condition.
\section{Random Networks}
Erd\"os-R\'enyi random networks of size $N$ are constructed where pairs
of nodes are connected with a probability $p$.
Figure \ref{Fig_ER_Stat} plots the statistical properties of networks with $0$, $1$ 
and $1-1/p_{\mathrm in}$ entries. 
The data is fitted with the Gaussian and GEV distributions (Eq.~\ref{eq_gev}). 
Figure~\ref{Fig_ER_Stat} is plotted for various values of $p_{\mathrm in}$ while keeping other network parameters 
the same.
The nature of the distribution is normal for
$p_{\mathrm in}$ = 0. 
As inhibitory connections are introduced, thereby inducing directionality into the
underlying network, the
complex eigenvalues start appearing in conjugate pairs. The
statistics of $R_{\mathrm max}$
are deformed as compared to that of the undirected network, which can not be characterized by any well-known 
statistics for regime
$0 \lesssim p_{\mathrm in} \lesssim 0.40$. 
For values of $p_{\mathrm in}$ lying between $0.40$ and $0.50$,
the statistics has an $\xi$ parameter value
close to the zero, indicating a convergence to the Gumbel distribution. 
Calculations of the shape parameter and detailed discussions
on fitting have been exemplified in \cite{supp}.
\begin{figure}[t]
\centerline{\includegraphics[width=0.9\columnwidth,height=4cm]{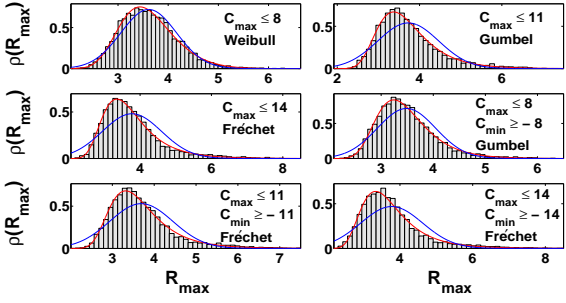}}
\caption{(Colour online) Statistics of $R_{\mathrm max}$ at different values
of $C_{max}$ and $C_{\mathrm min}$ for $p_{\mathrm in} = 0.5$.
The histograms are numerical results; blue and red lines correspond to the normal
and GEV fit, respectively. For each case $N=50$ and $<k>=15$.
All plots are obtained for 5000 realizations of the network in that ensemble.}
\label{Fig2}
\end{figure}
\begin{figure}[H]
\centerline{\includegraphics[width=0.9\columnwidth,height=4cm]{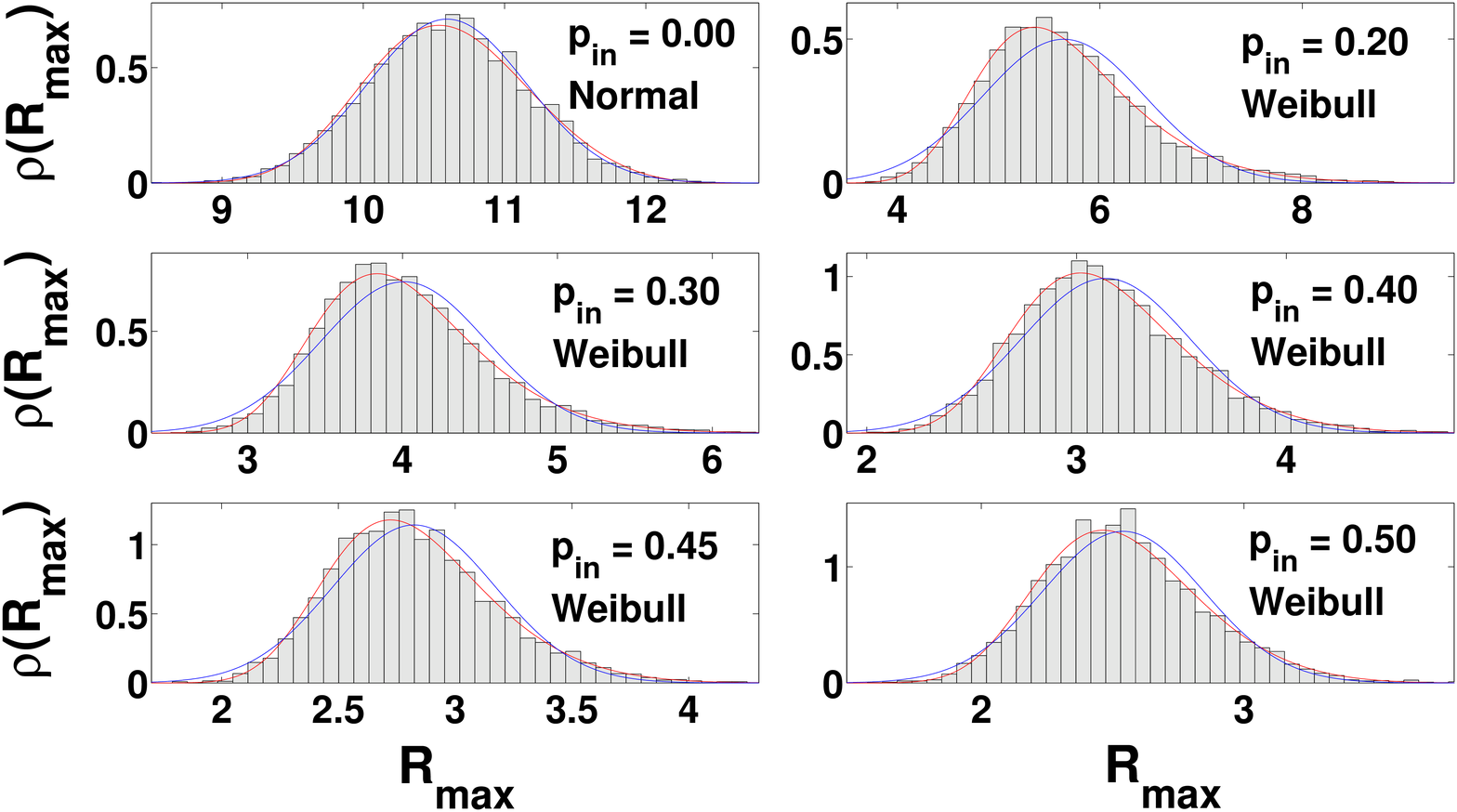}}
\caption{(Colour online) Statistics of $R_{\mathrm max}$ at different values
of $p_{\mathrm in}$ for Erd\"os-R\'enyi
networks with a strictly balanced condition.
The histograms are numerical results; blue and red lines correspond to normal
and GEV fit respectively. For each case, the statistics displays the Weibull distribution,
except $p_{\mathrm in}$ = 0
All plots are obtained for 5000 realizations of matrices with size 50 and $\langle k \rangle$ = 10.}
\label{Fig3}
\end{figure}
The behaviour of the column sum statistics provides an understanding of
the impact of network structure on the shape parameter of a GEV distribution. For the
balanced condition,
the mean and variance of the column sum are zero and
$Np(1/p_{\mathrm in} - 1)$ respectively. The maximum and minimum values of the
column sum for a particular matrix in the ensemble are denoted by $C_{\mathrm max}$ and $C_{\mathrm min}$. 
The Weibull, Gumbel, and Fr\'echet distributions are observed in Fig.\ref{Fig2}(a), (b) and
(c), respectively 
for an ensemble consisting of realizations of matrices generated by imposing
three different restrictions on $C_{\mathrm max}$ by keeping $p_{\mathrm in}$, $p$
and $N$ the same. An additional limitation on $C_{\mathrm min}$ to a particular value
shifts the shape parameter $\xi$ towards a positive value yielding Gumbel and Fr\'echet statistics
as illustrated by Fig.\ref{Fig2} (d), (e) and (f). The implications of restricting $C_{\mathrm min}$
and $C_{\mathrm max}$ are that they characterize the deviation from 
the strictly balanced condition, and interestingly, decide the shape parameter
of the statistics regardless of the denseness of underlying networks. 
\begin{figure}[t]
\centerline{\includegraphics[width=0.65\columnwidth,height=3.5cm]{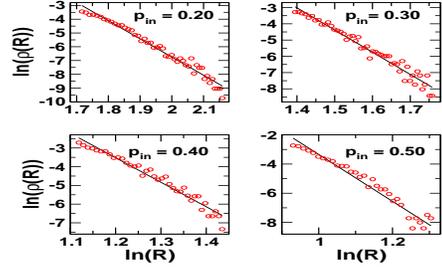}}
\caption{(Colour online) The tail behaviour of the real part of the eigenvalues at different values
of $p_{\mathrm in}$ for Erd\"os-R\'enyi networks with the strictly balanced condition.
For each case, $N=50$ and $\langle k \rangle=10$. }
\label{Fig4}
\end{figure}
Note that for the lower $p_{\mathrm in}$ values, the network has a significant
number of nodes connected to only excitatory nodes, thus failing to 
satisfy the strictly balanced condition.
In order to avoid this situation, only those realizations are chosen that lead to columns 
with at least one negative entry. Fig.~\ref{Fig3} depicts the statistics under 
the strictly balanced condition for a network of size $N=50$ and $p = 0.20$. The Weibull distribution is observed 
in the regime $0.20 \lesssim p_{\mathrm in} \lesssim 0.50$.
The statistics witness a 
sharp transition from the Gaussian to the Weibull at $p_{\mathrm in} = 0.2$. The estimated parameters of 
both the types of statistics and the detailed information of fitting
are addressed in \cite{supp}.

\subsection{Tail behaviour}
The nature of extreme value distribution can further be explained by the tail behaviour 
of the parent distribution\cite{book_gev}. In the case of the Weibull distribution the tail of the 
parent distribution follows a power-law with bounded maxima. Fig.~\ref{Fig4} 
plots tail behaviour extracted from the real part of the eigenvalues for the matrices 
associated with the Erd\"os-R\'enyi networks at different values of  
$p_{\mathrm in}$, which confirms the Weibull statistics 
as expected from the extremal eigenvalues
of this ensemble.
\subsection{Random matrices}
\begin{figure}[t]
\centerline{\includegraphics[width=0.9\columnwidth,height=4cm]{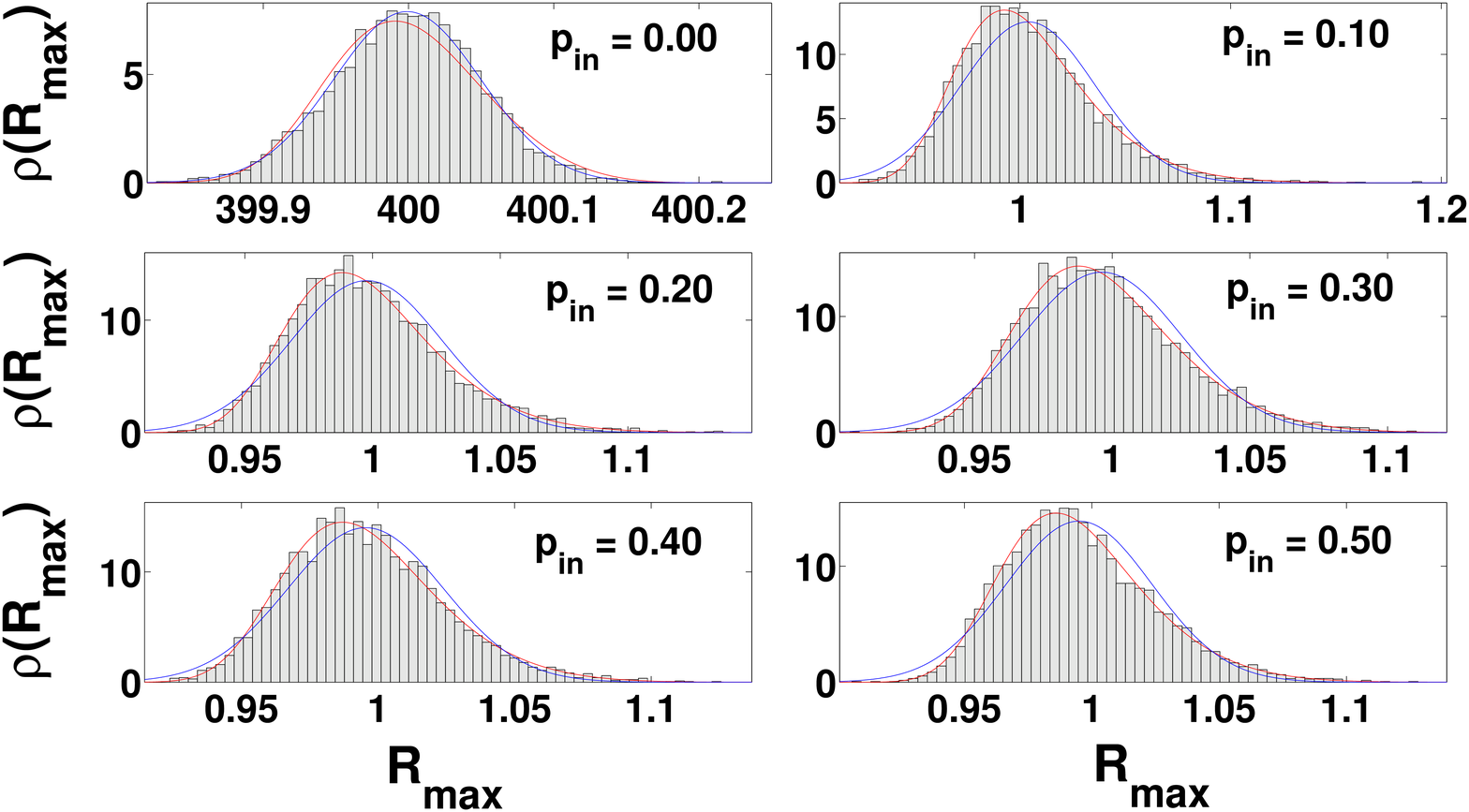}}
\caption{(Colour online) Statistics of $R_{\mathrm max}$ at different values
of $p_{\mathrm in}$ for random matrices. The histograms are
numerical results; blue and red lines correspond to normal
and GEV fit, respectively. For each cases the statistics display the Weibull distribution,
except $p_{\mathrm in}=0$ for which normal distribution is observed.
All plots are obtained for 5000 realizations of matrices with $N=400$.}
\label{RandomMatricesN400}
\end{figure}

Fig.~\ref{RandomMatricesN400} plots $\rho(R_{\mathrm max})$ for random matrices generated
using Gaussian distributed random numbers under the strictly balanced condition. This matrix represents the case of when coupling weights of 
inhibitory and excitatory connections are taken
from Gaussian distributed random numbers with mean $1-1/p_{\mathrm in}$
and 1, and standard deviation 0.05, as considered in the Ref.\cite{Rajan}.
The nature of the $R_{\mathrm max}$ distribution remains normal for $p_{\mathrm in}$ = 0,
whereas the Weibull distribution is observed for 
$0.10 \lesssim p_{\mathrm in} \lesssim 0.50$. The robustness of the Weibull
statistics in this parameter regime is 
indicated by a fixed value for the $\xi$ parameter. The mean and variance of 
the data remains constant for the strictly balanced condition in 
the range $0.10 \le p_{\mathrm in} \le 0.50$. 
\begin{figure}[H]
\centerline{\includegraphics[width=0.9\columnwidth,height=4cm]{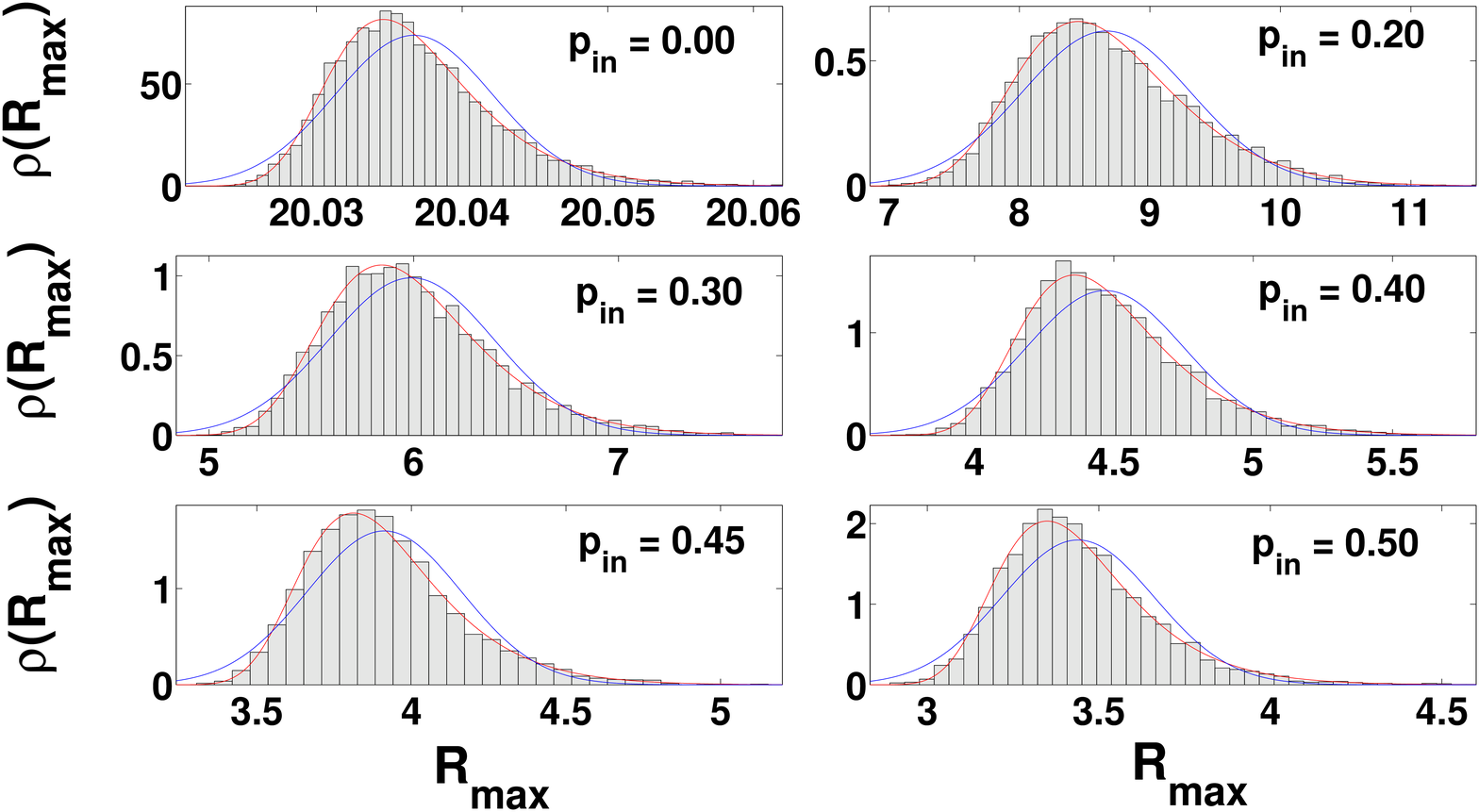}}
\caption{(Colour online) Statistics of $R_{\mathrm max}$ at different values
$p_{\mathrm in}$ for small-world networks with the strictly balanced condition.
The histograms are numerical results; blue and red lines correspond to normal
and GEV fit, respectively. For each case, the statistics exhibit the Weibull distribution.
The rewiring probability is fixed at $p_{\mathrm r}$ = 0.0256 characterizing small-world
transition.
All plots are obtained for 5000 realizations of matrices with $N=500$ and $\langle k \rangle=20$.}
\label{SwExtremeAv20N500}
\end{figure}
This result confirms the robustness of the Weibull distribution for the strictly 
balanced condition, as it leads to this distribution
independent of whether the matrix is modelled using a
random network, i.e. entries being 0 and 1, or
a random matrix, where entries are the Gaussian distributed random numbers.
The left panel of the phase diagram in Fig.~\ref{PhageSWregion}
illustrates this behaviour for various values
of $p$ and $p_{\mathrm in}$ for Erd\"os-R\'enyi networks under the strictly balanced condition. 
Very small values of $p_{\mathrm in}$ may yield a situation in which some
columns have only positive entries, and which do not allow the strictly balanced condition 
to be imposed, thereby making these values of $p_{\mathrm in}$ out of the scope
of the present study.

For some cases, the KS test accepts the normal as well as the Weibull distribution.
This happens because a particular shape parameter range, the Weibull distribution complies closely with 
the normal distribution \cite{Dubey}. 
For very high values of $p$, the conformation space of a network's structure is reduced,
which results in a lack of variation in network topology for an
ensemble. This might be a reason for the undefined shape of the statistics for
higher $p$ values. 
Model systems having a lower average degree
could be modelled by the Weibull distribution.
which might be due to many configuration possibilities in
the ensemble leading to a more fluctuations in the $R_{\mathrm max}$, leading
to a smooth shape of the statistics. 
The fact that most of the real-world networks are sparse \cite{sparse},
implies that they too can be modelled by this ensemble, which exhibits the Weibull distribution.
We further analyze the effects of different network configurations on the statistics of extremal 
eigenvalues under the strictly balanced condition.  
\begin{figure}
\centerline{
\includegraphics[width=.45\columnwidth,height=2.7cm]{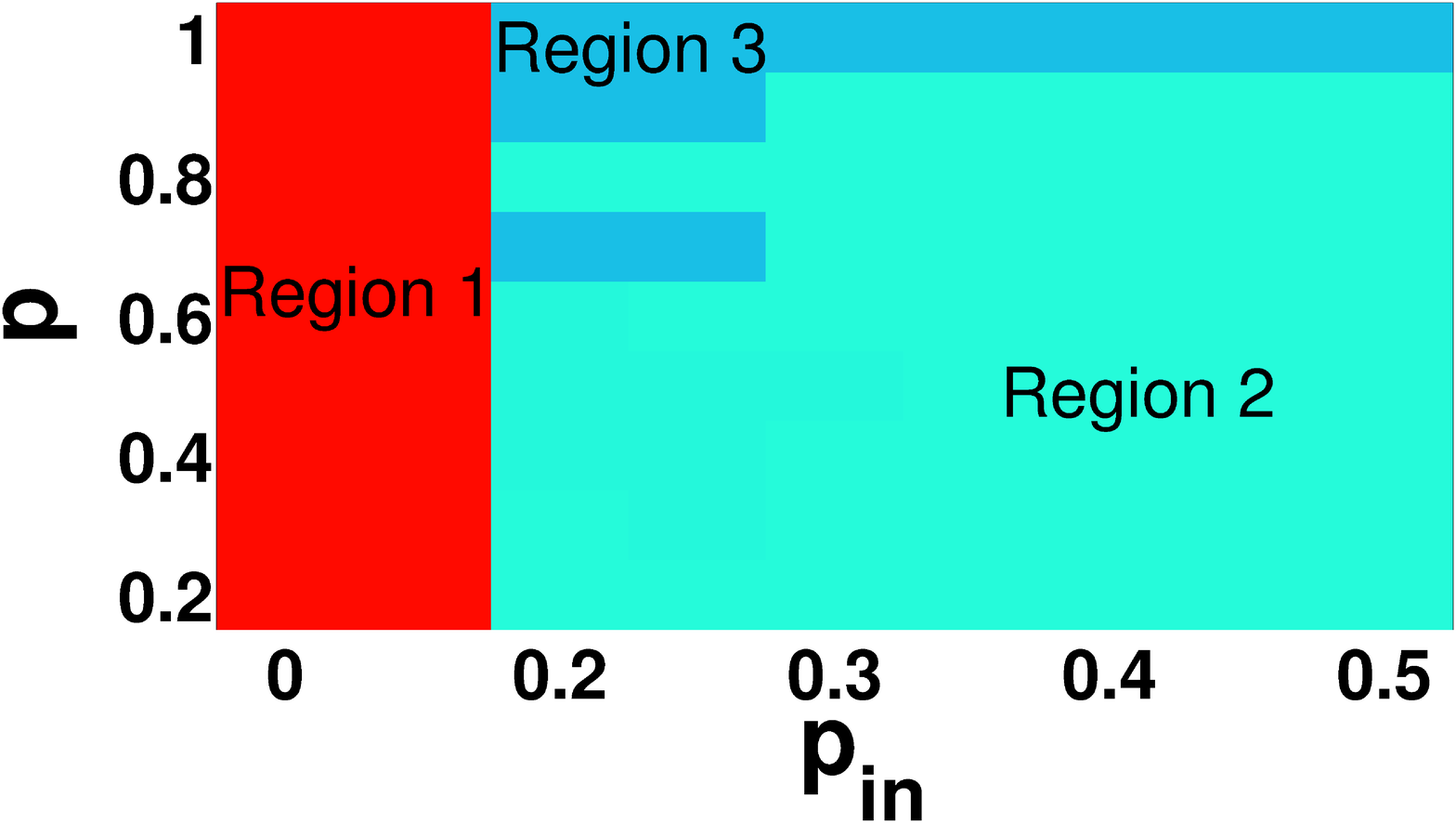}
\includegraphics[width=0.45\columnwidth,height=2.7cm]{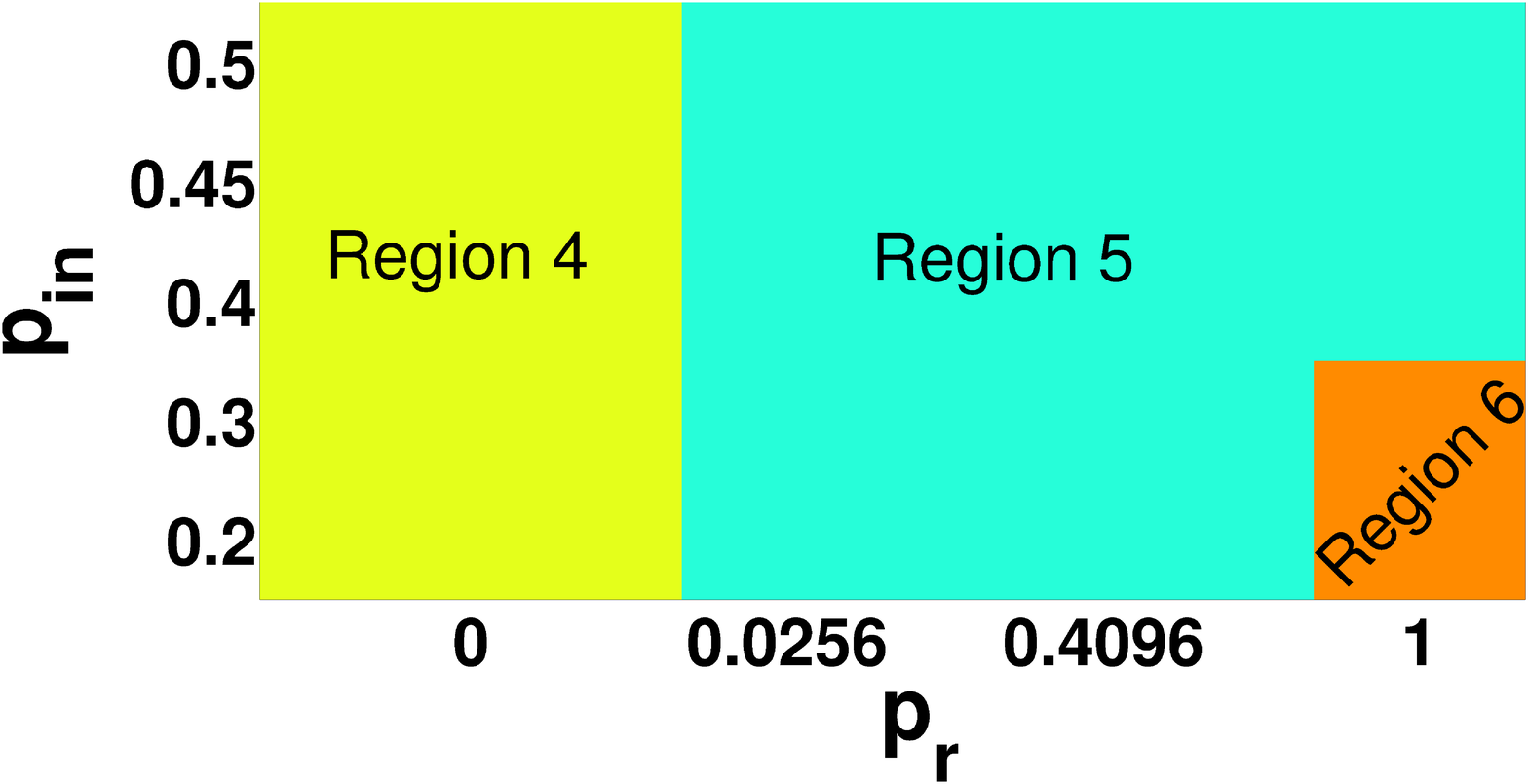}}
\caption{(Left panel) Statistical behaviour of $R_{\mathrm max}$ for Erd\"os-R\'enyi networks with 
$N=50$ at different values of $p_{\mathrm in}$ and $p$. Region 1 denotes the parameter region for which the
strictly balanced 
condition cannot be defined, because some columns have only non-negative entries. 
Region 2 corresponds to the distribution that is Weibull or close to Weibull. Region 3 stands for 
the undefined distribution.
(Right panel) For small-world networks with $N=500$ and $\langle k \rangle=20$ at different values of $p_{\mathrm in}$ and rewiring 
probability $p_{\mathrm r}$. Regions 4, 5 and 6 
represent Gumbel (or close to it), Weibull 
and normal distributions, respectively. 
All plots are obtained for 5000 realizations of the networks.}
\label{PhageSWregion}
\end{figure}
\begin{figure}
\centerline{\includegraphics[width=0.9\columnwidth,height=2.2cm]{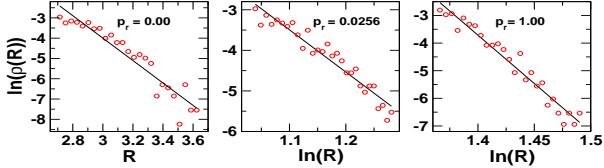}}
\caption{(Colour online) The tail behaviour of the real part of the eigenvalues at various values
of  $p_{\mathrm r}$ for small-world networks under the strictly balanced condition.
For each case, $N=100$ and $p_{\mathrm in}=0.50$.}
\label{tailSW}
\end{figure}   
\section{Small-World Networks}
First we consider small-world networks generated using the Watts-Strogatz algorithm.
Properties of many real-world networks, including the brain, are  prescribed by this small-world 
model \cite{Watts}. This type of network maintains the clustering coefficient close to that of the regular 
lattices, whereas the average diameter is close to that of the random networks. Small-world networks have been found in C. elegans, cat cortex, and macaque 
cortex and it has been shown that the efficiency of the brain to rapidly integrate information 
from both locally and distantly specialized brain areas increases with the
organization of small-world topology \cite{sparse}.

We generate small-world networks using the Watts-Strogatz model \cite{Watts}.
For $N=500$ and $\langle k \rangle=20$,
the rewiring probability is chosen as $p_{\mathrm r}$ = 0.0256, which 
corresponds to the small-world transition.
Fig.~\ref{SwExtremeAv20N500} confirms that
for all the $p_{\mathrm in}$ values, the statistics remains Weibull. All the
plots of Fig.\ref{SwExtremeAv20N500} except that which corresponds to 
$p_{\mathrm in}$ = 0.0, satisfy the strictly balanced 
condition.
The mean and variance of the $R_{\mathrm max}$ decrease monotonically with $p_{\mathrm in}$
for the strictly balanced condition.
However, the shape parameter ($\xi$) is most negative for $p_{\mathrm in}=0.20$, which 
corresponds closely to a real brain situation \cite{DiverseInhibitory}, reflecting 
a less right-skewed Weibull statistics. 
Information pertaining to the parameters estimated in the statistics are referred in \cite{supp}.

\subsection{Phase diagram for small-world networks}   
The phase diagram in Fig.~\ref{PhageSWregion} (right panel) demonstrates the behaviour of 
the statistics for various values
of $p_{\mathrm r}$ and $p_{\mathrm in}$. For $p_{\mathrm r} = 0$, only 
inhibitory couplings contribute to the statistics, and as a result the statistics  
of $R_{\mathrm max}$ is found close to the 
Gumbel. 
The tail behaviour displays an exponential decay, thereby
supporting the observed Gumbel distribution (Fig.~\ref{tailSW}).
As $p_{\mathrm r}$ increases, the contribution of 
structural variation also increases yielding more variation in the 
$R_{\mathrm max}$ statistics. Because the nature of the statistics is
determined by the shape parameter $\xi$, for a fixed
value of disorder ($p_{\mathrm r}$), the occurrence of all the three statistics are possible.
Increased disorder in network structure leads to an enhancement of the
value of the $\sigma$ parameter (Eq.\ref{eq_gev}).
Interestingly, at the small-world transition, the statistics display a 
transition from the Gumbel to the Weibull. 
At the small-world transition,
the underlying network has sufficient randomness \cite{SJ_EPL2009},
which might be one of the reasons behind a drastic change in spectral behaviour
from the exponential. This indicates that the small-world 
transition appears as a critical point in terms of the stability of 
the underlying network in that ensemble.

The results for Erd\"os-R\'enyi networks differ slightly from
the networks generated using the small-world algorithm at $p_{\mathrm r}=$ 1. 
For lesser $p_{\mathrm in}$ values, instead of the Weibull, the normal distribution is observed.
This might be due to the fixed total number of degrees occurring
in all the realizations for networks generated using the small-world model. 
Note that for Erd\"os-R\'enyi networks,
at a particular $p$ value
there exists a fluctuation in the total degree for the different network
realizations.

\section{Scalefree Networks}
In this section we present results for the scalefree network architecture, generated 
\begin{figure}[t]
\centerline{\includegraphics[width=0.9\columnwidth,height=4cm]{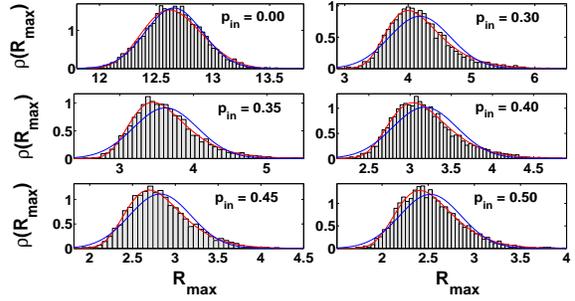}}
\caption{(Colour online) Statistics of $R_{\mathrm max}$ for different values
of $p_{\mathrm in}$ for scalefree networks with the strictly balanced condition. 
The histograms are numerical results; blue and red lines correspond to normal
and GEV fit, respectively. For each case, the statistics show the Weibull distribution, 
except $p_{\mathrm in}$ = 0. 
All plots are obtained for 5000 realizations of networks with $N=100$ and $\langle k \rangle=8$.}
\label{SFAv8st_ba}
\end{figure} 
using the preferential growth algorithm \cite{Albert}. After introducing inhibitory 
connections with the probability $p_{\mathrm in}$, the strictly balanced
condition is imposed.
Fig.~\ref{SFAv8st_ba} demonstrates that for all 
$p_{\mathrm in}$ values, the statistics remain Weibull. However, the KS test accepts the 
normal distribution as well for $p_{\mathrm in}$ =0. The mean and variance of the data 
remains constant for the strictly balanced condition and for
the different $p_{\mathrm in}$ values (i.e.  0.30, 0.35, 0.40, 0.45 and 0.50). 
The reason for discarding cases with $p_{\mathrm in} \le 0.20$ is that these values
do not yield enough realizations that satisfy the strictly balanced condition. 
It happens due to the presence of a large number of nodes having a lesser degree, 
as compared to the Erd\"os-R\'enyi networks. The observed statistics can further be explained from the
tail behaviour of the parent distributions. Fig.~\ref{tailSF} displays 
a consistent power-law behaviour with the increase in
$\langle k \rangle$, whereas the shape
parameter of GEV monotonically decreases with an increase
in $p_{\mathrm in}$. The estimated parameters information is 
referred to \cite{supp}.
\begin{figure}[t]
\centerline{\includegraphics[width=0.9\columnwidth,height=2.2cm]{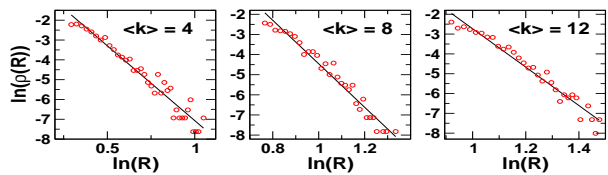}}
\caption{(Colour online) The tail behaviour of the real part of the eigenvalues at different values
of $\langle k \rangle$ for scalefree networks with the strictly balanced condition.
For each case, $N=100$ and $p_{\mathrm in}=0.50$.}
\label{tailSF}
\end{figure}   

\section{Discussions and Conclusion}
The nature of extreme values distribution for many real-world systems 
is associated with various shape parameters.
The right skewness reflects the chances of occurrence of higher values. 
The effect of the strictly balanced condition dominates the behaviour of
extreme value statistics, 
in particular to a fixed Weibull statistics. This condition is so strong that
even changes in the interaction patterns do not affect the distribution behaviour.

Origin of the Weibull distribution for the strictly balanced condition could be explained
by the fact that the strictly balanced condition shifts
the outliers into the bulk of spectra,
i.e. $R_{\mathrm max}$ becomes bounded \cite{Rajan}.
The observed Weibull statistics is supported further by the tail behaviour of the parent 
distribution which follows a power-law with bounded maxima. 

The strictly balanced condition yields the Weibull distribution for networks
with structural variations such as Erd\"os-R\'enyi random networks and
scalefree networks. The 1-d lattice
structure lacks any structural variation in the ensemble leading to a deviation from
the Weibull distribution even for the strictly balanced condition.  
We demonstrated that at the small-world transition, a network has sufficient
structural variations or {\it randomness} leading to a
less right-skewed statistics governed by the
Weibull distribution, consequently making the system more stable.

The Weibull distribution does not display any significant change with the change
in the average degree of the 
network in the balanced condition, whereas previous work \cite{Extreme} 
demonstrates a deviation from the Weibull to the Fr\'echet 
distribution via Gumbel as connectivity of the network increases by keeping
$p_{\mathrm in}$ fixed at $0.5$.
The reasons for the networks with a lower $p$ value (corresponding to a lower average degree)
following the Weibull distribution is that
such matrices have fewer fluctuations around the strictly balanced condition and
exhibit similar statistics to that observed for the strictly balanced condition. 
However, higher values of 
$p$ yield matrices with more deviations than those satisfying
the strictly balanced condition, and as a result lead to an increased number
of outliers from the bulk part of the eigenvalues, consequently
resulting in a transition from the Weibull statistics. 

The extreme value theory might enhance our understanding of stability properties of
real brain systems. 
For instance, model
networks capturing real brain network properties, such as small-world architecture and
a 20-80\% inhibitory to excitatory ratio, tend to witness fewer right-skewed $R_{\mathrm max}$
statistics compared to other possible 
values of $p_{\mathrm in}$. The higher values of $R_{\mathrm max}$ are more 
likely to generate right-skewed statistics with higher variances. The variances can be managed
by weight scaling the connections. 
The combined framework of the network
architecture and the strictly balanced situation thus emulates the existence of stable statistics
upon capturing realistic brain scenario.
Future studies will incorporate other network architectures, particularly those
having community structures \cite{community}.  Recently the stability of eco-systems has been analyzed using 
the spectral properties of 
underlying matrices \cite{Allesina}. The framework presented in this Letter
can be extended to have a proper understanding 
of other such complex systems \cite{unpublished}. 

\section{Acknowledgment} SJ thanks the DST and CSIR for funding. 
SKD acknowledges the University grants commission, India for
financial support. 

\end{document}